\documentstyle[aps,preprint]{revtex}
\begin{document}
\draft
\date{\today}
\title
{Magnetic field of an in-plane vortex outside a layered superconductor}
\author{J.R. Kirtley}
\address{ IBM T.J. Watson Research Center, P.O. Box 218, Yorktown Heights,
NY
10598}
\author{V. G. Kogan and J. R. Clem }
\address{ Ames Laboratory and Physics Department ISU, Ames, IA
50011}
\author{K.A. Moler}
\address{ Dept. of Applied Physics, Stanford University, Stanford, CA
94305}
\maketitle

\begin{abstract}
We present the solution to London's equations
for the magnetic fields of a vortex
oriented parallel to the planes, and normal to a crystal face,  of a
layered superconductor. These expressions account for
flux spreading at the superconducting surface, which can change the
apparent
size of the vortex along the planes by as much as 30\%. We
compare these expressions with experimental results.
\end{abstract}
\narrowtext
\subsection{Introduction}
Recently, scanning SQUID microscope magnetic imaging
of interlayer vortices trapped between the planes of layered
superconductors
has been used to make
direct measurements of the interlayer penetration
depth in several layered superconductors\onlinecite{kamsci},
\onlinecite{vdmnat},\onlinecite{hgprl},\onlinecite{orgjpc}.
These experiments
provide local measurements
of the interlayer supercurrent density, which have implications
for the validity of the
interlayer tunneling model\onlinecite{ilt} as a candidate mechanism
for superconductivity in the
high critical temperature cuprate superconductors.

To date the quantitative modelling of these experiments has
assumed that the vortex fields at the superconductor-vacuum
interface are the same as those in the bulk, neglecting the well known
effect that the magnetic fields from vortices spread as they
approach the superconductor-vacuum surface from within the
superconductor.
Exact theoretical expressions
exist for a vortex in an isotropic London's model\onlinecite{pearl}.
For a vortex oriented perpendicular
to the surface in a superconductor with an isotropic penetration depth
$\lambda$, the fields above the surface can be approximated
by a magnetic monopole located a distance $\lambda$ below the surface
\onlinecite{pearl},\onlinecite{hess}.
This means that the spatial extent of the
magnetic fields at
the surface is larger than in the bulk of the superconductor. If
the bulk expressions were used to fit data at the surface, the fitted
value of the penetration depth would be longer than the real value.
This effect must be accounted
for in making quantitative  estimates of the penetration depths by
magnetic imaging measurements. For this purpose it is useful to
examine vortex spreading at the surface for a highly anisotropic
superconductor, since
recent experiments have studied vortices in superconductors with
$\lambda_c/\lambda_{ab} \sim$ 10-100.

It is well known that the anisotropic London model is
appropriate for describing a stack of Josephson-coupled superconducting
layers at length scales large compared to the interlayer spacing.
In this paper we present an exact solution of London's equations for
a straight vortex approaching a superconductor-vacuum interface normal
to the interface, in an anisotropic superconductor. We show how flux
spreading near this interface effects the magnetic fields above the
interface, and show that there is good agreement between these
theoretical results and scanning SQUID microscope
measurements on single crystals of the layered high-T$_c$ cuprate
superconductor Tl$_2$Ba$_2$CuO$_{6+\delta}$ (Tl-2201)\onlinecite{kamsci}.

\subsection{The model}

A method for finding the field distribution of a straight vortex crossing a
plane surface of an anisotropic superconductor has been developed in Ref.
\onlinecite{KSL}. We will outline this method
and apply it to the case of a
vortex, oriented along $b$ in the $ab$ plane of a uniaxial material,
which
crosses the plane face $ca$ of the crystal. For a vortex not too close to
the
crystal corners, the crystal surface $ca$ can be taken as an
infinite plane. We choose the
coordinates
$x,y,z$ corresponding to $c,a,b$ of the crystal as shown
in Fig. 1. Then the mass tensor is diagonal: $m_{xx}=m_3,
m_{yy}=m_{zz}=m_1$. The standard normalization $m_1^2m_3=1$ is implied.
The method consists of solving London's equations for the field
inside the superconductor and matching the result to a solution of
Maxwell's equations in the vacuum outside the sample.
For the isotropic case, the problem is simplified by the cylindrical
symmetry of the field distribution\cite{pearl,clem}. This is not
the case for anisotropic materials, and a more general approach is
needed.

Inside the superconductor, the field ${\bf h}({\bf r},z)$, with ${\bf
r}=\{x,y\}$, satisfies the London equations \onlinecite{K81}:
\begin{equation}
h_i-{4\pi\over c}\lambda^2m_{kl}e_{ils}\frac{\partial j_k}{\partial x_s}=
\phi_0 \delta({\bf r})\delta_{iz}\,.
\label{London}
\end{equation}
Here, ${\bf j}$ is the current density, $\phi_0 = hc/2e$ is
the superconducting flux quantum, and
the average penetration depth
$\lambda=(\lambda_{ab}^2\lambda_c)^{1/3}$
($\lambda_{ab}^2=m_1\lambda^2$, $\lambda_c^2=m_3\lambda^2$).

Deep inside the superconductor,
the field  ${\bf h}({\bf r})$ has only a $z$ component.
However, near the exit from the sample at $z=0$, the vortex ``opens up"
and
$h_x,h_y$ are no longer zero. In other words, Eq. (\ref{London}) is a
system
of three linear differential equations for $h_x, h_y$, and $h_z$ with a
non-zero right-hand side (RHS). The general solution is then
\begin{equation}
{\bf h}={\bf h}^{(0)}+{\bf h}^{(v)}\,,
\label{sum}
\end{equation}
 where ${\bf h}^{(0)}$ solves the homogeneous system with zero RHS, whereas
${\bf h}^{(v)}$ is a particular solution of the full system (\ref{London}).
The latter can be taken as the field of an infinitely long unperturbed
vortex
along $z$; this assures correct singular behavior at the vortex axis. The
Fourier transform of this field is
\begin{equation}
{\bf h}^{(v)}=\frac{\phi_0}{1+\lambda_{ab}^2k_x^2+\lambda_c^2k_y^2}\, {\hat
{\bf z}}\,.
\label{h_v}
\end{equation}
With this choice of ${\bf h}^{(v)}$, the field ${\bf h}^{(0)}$  is the
correction due to the surface of the unperturbed vortex field ${\bf
h}^{(v)}$.
We note that the Clem-Coffey result for a vortex parallel to the layers
of a Josephson coupled layered superconductor
reduces to ${\bf h}^{(v)}$ if
one disregards the core correction\onlinecite{cc}.

Because the only sample boundary is parallel to the plane $xy$, we Fourier
transform Eq. (\ref{London}) with respect to $x,y$. We are then left with
the system of equations for
${\bf h}({\bf k},z)=\int d{\bf r}\exp(-i{\bf k}\cdot{\bf
r}){\bf h}({\bf r},z)$:
\FL
\begin{eqnarray}
m_1h_x^{\prime\prime}-(1+m_1k_y^2)h_x+m_1k_xk_yh_y-im_1k_xh_z^{\prime}=0\,,
\nonumber \\
m_1k_xk_yh_x+m_3h_y^{\prime\prime}-(1+m_1k_x^2)h_y-im_3k_yh_z^{\prime}=0\,,
\label{L2} \\
im_1k_xh_x^{\prime}+im_3k_yh_y^{\prime}+(1+m_1k_x^2+m_3k_y^2)h_z=\phi_0\,.
\nonumber
\end{eqnarray}
For brevity, we have set the average $\lambda$ as the unit of length so
that
$\lambda_{ab}^2=\lambda^2m_1$ and $\lambda_c^2=\lambda^2m_3$ are replaced
with
$m_1$ and $m_3$; the prime in the above equations denotes $d/dz$. The field
${\bf h}^{(0)}({\bf k},z)$ satisfies the {\it homogeneous} system of {\it
linear second order ordinary differential} (with respect to the variable
$z$) equations, i.e., it is a linear combination of exponential functions
of
$z$:
\begin{equation}
{\bf h}^{(0)}({\bf k},z) = \sum_n {\bf H}^{(n)} e^{\alpha_nz}\,.
\label{h-0}
\end{equation}
The $z$ independent coefficients ${\bf H}^{(n)}({\bf k})$ and
$\alpha_n({\bf
k})$ are still to be determined. Each term in the sum (\ref{h-0}) should
satisfy  separately the system (\ref{L2}) with zero RHS.
Omitting the
label $n$ we write this system as
\begin{equation}
\Delta_{ij}H_j=0
\label{homog}
\end{equation}
with a symmetric matrix $\Delta_{ij}$:
\begin{eqnarray}
\Delta_{xx}&=&1+m_1k_y^2-m_1\alpha^2\,,\nonumber\\
\Delta_{xy}&=&-m_1k_xk_y \,,\nonumber\\
\Delta_{xz}&=&i\,m_1k_x\alpha\,,\label{Delta}\\
\Delta_{yy}&=&1+m_1k_x^2-m_3\alpha^2\,,\nonumber\\
\Delta_{yz}&=&i\,m_3k_y\alpha\,,\nonumber\\
\Delta_{zz}&=&1+m_1k_x^2+m_3k_y^2\,.\nonumber
\end{eqnarray}
The determinant of this matrix must be zero, which provides all possible
values
of $\alpha$:
\begin{eqnarray}
\alpha_{1,2}&=&\pm\Big(\frac{1+m_1k^2}{m_1}\Big)^{1/2},\label{al1}\\
\alpha_{3,4}&=&\pm\Big(\frac{1+m_1k_x^2+m_3k_y^2}{m_3}\Big)^{1/2}.\label{al3}
\end{eqnarray}
Deep inside the superconductor, the surface correction ${\bf
h}^{(0)}(z\rightarrow -\infty)$ must vanish, implying
$\alpha_1$ and $\alpha_3$ must be positive. The
homogeneous
system (\ref{homog}) allows one to express (for each of these
$\alpha$'s)  two out of three components $H_i$ in terms of the
third. We obtain after simple algebra:
\begin{eqnarray}
H_x^{(1)}&=&i\,\frac{1+m_1k_x^2}{m_1k_x\alpha_1}\,H_z^{(1)}\,,\quad
H_y^{(1)}=i\,\frac{k_y}{\alpha_1}\,H_z^{(1)}\,;
\label{H1}\\
 H_x^{(3)}&=& 0\,,\qquad\qquad\qquad\quad
H_y^{(3)}=i\,\frac{\alpha_3}{k_y}\,H_z¶{(3)}\,.
\label{H3}
\end{eqnarray}
Thus, the
field inside the sample will be determined completely after $H_z^{(1)}$ and
$H_z^{(3)}$ are found from the boundary conditions at the sample surface.

The field outside the sample is described by div${\bf h }=0$ and curl${\bf
h}=0$, so that one looks for ${\bf h}=\nabla\varphi$ with
$\nabla^2\varphi=0$.
The general solution of Laplace's equation which vanishes at
$z\rightarrow\infty$ is
\begin{equation}
\varphi ({\bf r},z)=\int\frac{d^2{\bf k}}{(2\pi)^2}\,\varphi({\bf
k})\,e^{i{\bf
k}\cdot{\bf r}-kz}\,.
\label{Lapl}
\end{equation}
The 2D Fourier transform is defined by
\begin{equation}
\varphi ({\bf k})=e^{kz}\int d^2{\bf r}\,\varphi({\bf r},z)\,e^{-i{\bf
k}\cdot{\bf r}}\,.
\label{invLapl}
\end{equation}

The boundary conditions at the  free surface $z=0$ consist of continuity
of the three field components:
\begin{eqnarray}
ik_x\,\varphi&=&H_x^{(1)}+H_x^{(3)}\,,\nonumber\\
ik_y\,\varphi&=&H_y^{(1)}+H_y^{(3)}\,,\label{bc}\\
-k\,\varphi&=&h^{(v)}_z+H_z^{(1)}+H_z^{(3)}\,.\nonumber
\end{eqnarray}
The components $H_{x,y}^{(1,3)}$ are expressed in terms of $H_z^{(1,3)}$ in
Eqs. (\ref{H1}), (\ref{H3}), so that the system (\ref{bc}) can be
solved to find $\varphi$ along with all $H_i$'s. We are interested here
primarily in the field outside the sample:
\begin{equation}
\varphi ({\bf k})= -
\frac{\phi_0\,(1+m_1k_x^2)}{m_3\alpha_3\,[m_1k_x^2\,\alpha_3(k+\alpha_1)
+k\alpha_3+k_y^2]}\,.
\label{phi}
\end{equation}
It is readily verified that for the isotropic material Eq.
(\ref{phi}) reduces to the known result by Pearl: $\varphi = -
 \phi_0 /\alpha_{is} k (k+\alpha_{is})$ where $\alpha_{is}$ is the
isotropic version of either $\alpha_1$ or $\alpha_3$ \onlinecite{pearl}.

  Since {\it outside} the sample ${\bf h}=\nabla\varphi$, we have
\begin{equation}
h_{x,y}({\bf k}) = ik_{x,y}\,\varphi ({\bf k})\,,\qquad
h_z({\bf k}) = -k\,\varphi ({\bf k})\,.\label{outside}
\end{equation}
 Then, for example, the
\begin{equation}
 h_z ({\bf r},z)=-\int\frac{d^2{\bf k}}{(2\pi)^2}\,k\varphi({\bf
k})\,e^{i{\bf
k}\cdot{\bf r}-kz}\,.
\label{h_z}
\end{equation}
In particular, the total magnetic
flux through any plane $z=z_0$ is given by $h_z(k=0,z_0)=\phi_0$ as
expected.

The field {\it inside} the sample is given by Eqs. (\ref{sum}),
(\ref{h_v}), (\ref{h-0}). The coefficients ${\bf H}^{(1,3)}$ in
\begin{equation}
{\bf h}^{(0)}({\bf k},z) =  {\bf H}^{(1)} e^{\alpha_1z}+{\bf H}^{(3)}
e^{\alpha_3z}\,
\end{equation}
are obtained by solving
(\ref{bc}), (\ref{H1}), and (\ref{H3}):
\begin{eqnarray}
{\bf H}^{(1)} &=&\varphi({\bf k})\Big\{ ik_x\,,\,\,
i\frac{m_1k_x^2k_y}{1+m_1k_x^2}\,,\,\,
\frac{m_1k_x^2\alpha_1}{1+m_1k_x^2}\Big \}\,,\label{H11} \\
 {\bf H}^{(3)} &=&\varphi({\bf k})\Big\{0\,,\,\,\frac{i
k_y}{1+m_1k_x^2}\,,\,\,\frac{k_y^2}{\alpha_3
(1+m_1k_x^2)} \Big\}\,.
 \label{H31}
 \end{eqnarray}
Thus, for example,
\begin{equation}
 h_x ({\bf r},z)=\int\frac{d^2{\bf k}}{(2\pi)^2}\,ik_x\,\varphi({\bf
k})\,e^{i{\bf k}\cdot{\bf r}+\alpha_1 z}\,.
\label{h_x}
\end{equation}

For what follows, we will only concern ourselves with the fields outside
of the superconductor.
In the experiment of Ref.\onlinecite{kamsci}, the component $h_z$ was
probed
with a SQUID pickup loop which was much larger than the penetration depth
$\lambda_{ab}$. One therefore expects the
instrument to measure a flux nearly equal to the pickup loop size times
\begin{equation}
{\cal H}_z(x,y)=\int_{-\infty}^{\infty}h_z(x,y,z)\,dx \,.
\label{F}
\end{equation}
The vortex spreads as it approaches from below
the superconducting surface in the
$x$-direction
as well as in the $y$-direction; nevertheless numerical estimates show that
under typical conditions the experimental signal is well represented by
 Eq. (\ref{F}).
 Then we obtain:
\begin{equation}
\frac{ {\cal H}_z(y,z)}{\phi_0}=\int_{-\infty}^{\infty}{dk_y\over
2\pi}\,\frac{e^{ik_yy-|k_y|z}}{ m_3\alpha\,( \alpha+|k_y|)}\,
\label{phi_y}
\end{equation}
 with $\alpha =\alpha_3(k_x=0)=(m_3^{-1}+k_y^2)^{1/2}$. In conventional
units,
\begin{equation}
\frac{ {\cal H}_z(y,z)}{\phi_0}= \int_{-\infty}^{\infty}{dk_y\over
2\pi}\,\frac{e^{ik_yy-|k_y|z}}{
 \alpha\,( \alpha+\lambda_c|k_y|)}\,
\label{conv.units}
\end{equation}
with  $\alpha = (1+\lambda_c^2k_y^2)^{1/2}$.  It is worth noting that the
quantity ${\cal H}_z(y,z)$ depends only on $\lambda_c$.

After the substitution $\lambda_ck_y=\sinh u$,  Eq. (\ref{conv.units})
takes the form
\FL
\begin{eqnarray}
\pi\lambda_c\frac{ {\cal H}_z(x,y)}{\phi_0}&=&
\int_0^{\infty}du\, e^{-u-z^{\prime}\sinh u}\,\cos
(y^{\prime}\sinh u)\nonumber\\
&=&{\rm Re} \int_0^{\infty}du\, e^{-u- w\sinh u}\nonumber\\
&=&{\rm Re} \Big[{1\over  w}+{\pi\over 2}\Big({\bf
E}_1( w)+Y_1( w)\Big)\Big]\,,
 \label{0-infty}
\end{eqnarray}
Here, $y^{\prime}=y/\lambda_c$, $z^{\prime}=z/\lambda_c$, and $w=
(z+iy)/\lambda_c$; ${\bf E}_1 $ and $Y_1 $ are  Weber's and  Neumann
functions,
 see Ref. \onlinecite{RG}.

   For $|w|\ll 1$ (both $y$ and $z$ are small relative to $\lambda_c$) we
have
\cite{RG}:
\FL
\begin{eqnarray}
&&\pi\lambda_c\frac{ {\cal H}_z(y,z)}{\phi_0}= {\rm Re}\Big[1+{w\over
2}\Big(\ln\frac{w}{2}+\gamma -{1\over 2}\Big)\Big] \label{small w}\\
&&= 1+{z\over
2\lambda_c}\Big(\ln\frac{\sqrt{y^2+z^2}}{2\lambda_c}+0.077\Big) -{y\over
2\lambda_c}\tan^{-1}{y\over z}\,\nonumber
\end{eqnarray}
($\gamma =0.577$ is Euler's constant).
If  $|w|\gg 1$ (at least one of $y$ or $z$ is large relative to
$\lambda_c$), we obtain:
\begin{eqnarray}
\pi\lambda_c\frac{ {\cal H}_z(y,z)}{\phi_0}&=&{\rm Re}\Big[{1\over w}
-{1\over w^2} +O(|w|^{-4})\Big]\nonumber \\
&=&{z\lambda_c\over
y^2+z^2}- \frac{ (z^2-y^2)\lambda_c^2}{(z^2+y^2)^2}+...\,. \label{large w}
\end{eqnarray}
At large distances from the vortex exit, the second term can be neglected
and
$\lambda_c$ drops out of the result; this is expected since the field
there is
approaching the Coulomb form with no trace of material properties.

To form a complete picture
of the field distribution outside the sample, we
imagine that the same SQUID probe is oriented in the $xz$ plane so that the
$y$-component of the field (integrated over the probe area) is measured.
Then the SQUID flux
will be nearly equal to the pickup loop size times
\begin{eqnarray}
{\cal H}_y(y,z)&=&\int_{-\infty}^{\infty}h_y(x,y,z)\,dx \nonumber\\
&=&\int_{-\infty}^{\infty}{dk_y\over 2\pi}ik_y\varphi(0,k_y)
e^{ik_yy-€k_y€z}\,.
\label{Hy}
\end{eqnarray}

The two-dimensional field ${\vec{\cal H}}$ satisfies div${\vec{\cal
H}}$ = curl${\vec{\cal H}}=0$. Hence, it can be written as ${\vec{\cal
H}}=\nabla\Phi$ with $\nabla^2\Phi =0$. This implies that ${\cal
H}_z$ and ${\cal H}_y$ are real and imaginary parts of the same
analytic function given in Eq. (\ref{0-infty}):\cite{LL}
\begin{equation}
{\cal H}_y(y,z) =\frac{\phi_0}{\pi\lambda_c} {\rm Im} \Big\{{1\over
w}+{\pi\over 2}\Big[{\bf E}_1( w)+Y_1( w)\Big]\Big\}\,.
 \label{H_y}
\end{equation}
In particular, the asymptotic form of this function for $€w€\gg 1$ can
be written as
\begin{equation}
{1\over w}-{1\over w^2}\approx {1\over w+1}\,,
\end{equation}
which implies that at large distances the field  ${\vec{\cal H}}$
behaves as a field of a 2D "charge" situated at $z=-\lambda_c$.

\subsection{Results}

Figure 2(a) shows streamlines of ${\vec{\cal H}}(y,z)$
for a single anisotropic vortex centered at $x=0$, $y=0$.
These streamlines were generated numerically as follows:
the starting points of the lines were at $z/\lambda_c$=-3, with a
spacing in $y$ between the lines proportional to
$(\partial{\cal H}_z/\partial y)^{-1}$ at $z/\lambda_c=-3$. Small steps
$z=z+\Delta z$,
$y=y+\Delta y$, with
$\Delta z = \delta \sin\,\theta$, $\Delta y = \delta \cos\,\theta$,
were generated, with $\theta = \tan^{-1}({\cal H}_z(y,z)/{\cal H}_y(y,z))$.
The fields were recalculated at the new positions, and then the process
was repeated until $\mid z/\lambda_c \mid > 3$ or $\mid y/\lambda_c \mid >
3$.
Figure 2(b) shows the results if field spreading below the surface
is neglected:
\begin{equation}
b_z(x,y,z<0) = \frac{\phi_0}{2\pi \lambda_{ab}
\lambda_c} K_0(\tilde R),
\label{CK}
\end{equation}
where $K_0$ is a modified Bessel function of the second kind of order 0,
${\tilde R} = [(s/2\lambda_{ab})^2 +
(x/\lambda_{ab})^2 + (y/\lambda_{c})^2]^{1/2}$, and $s$ is
the interplanar spacing\onlinecite{cc}. For $s\ll\lambda_{ab}$, Eq.
(\ref{CK})
has the Fourier transform given in Eq. (\ref{h_v}). The fields for $z>0$
are
given by
\begin{equation}
b_z ({\bf r},z)=\int\frac{d^2{\bf k}}{(2\pi)^2}b_z({\bf
k})\,e^{i{\bf k}\cdot{\bf r}-k z}\,,
\end{equation}
where
\begin{equation}
b_z({\bf k})= \int d^2{\bf k}b_z(x,y,z=0) e^{-i{\bf k}\cdot{\bf r}}.
\end{equation}
\onlinecite{wikswo}. The fields in the $x$ and
$y$ directions are treated
similarly, using the relations $b_x({\bf k}) = -ik_xb_z({\bf k})/k$, and
$b_y({\bf k}) = -ik_yb_z({\bf k})/k$. The calculated fields were integrated
over $x$ and streamlines were generated just as for the  exact London
expressions. This is the procedure used
to model the experimental results in Ref.\onlinecite{kamsci}.
In both cases the fields extend into the vacuum nearly isotropically at
large distances, as if a point monopole source were placed near
$z=-\lambda_c$.
In the full treatment, Figure 2(a), the fields spread
as the vortex approaches the surface from inside the superconductor.

A one-dimensional rendering of our results above the surface of the
superconductor
is shown in Figure 3, which
plots $\pi \lambda_c {\cal H}_z(y,z)/\phi_0$ as a function of
$y/\lambda_c$ for
several values of $z/\lambda_c$. For comparison, the results neglecting
vortex spreading are also shown. Note that ${\cal H}_z(0,0)$, which
approximately
indicates the peak signal in an experiment,
is overestimated by a factor of $\pi/2$ if vortex spreading is
neglected. Also, the full width at half-maximum of the flux contour is
1.87$\lambda_c$ for the full theory, while it is 1.37$\lambda_c$ if flux
spreading is neglected. Therefore the neglect of flux spreading could
result in an overestimate of $\lambda_c$ by 30\%.

\subsection{Comparison with experiment}

Previous analyses of experimental data\onlinecite{kamsci},
\onlinecite{vdmnat}, neglected the
effect of vortex spreading at the surface. In retrospect, this neglect
was not unreasonable, given the quantitative agreement
between the interlayer coupling strength obtained from vortex
imaging measurements and from the
Josephson plasma resonance\onlinecite{vdmnat}.
With the full theory
presented in this paper, it is now possible to quantitatively
examine this assumption. However, we note that
there are additional systematic experimental uncertainties in
this technique.  These errors include the effect
of macroscopic screening currents which may create a slightly
inhomogeneous background, a relative angle
of 10-20 degrees between the surface of the superconductor and
the SQUID pickup loop, the uncertainty in
the exact value of the height of the pickup loop, and some effect
of the leads to the pickup loop on the
effective shape of the pickup loop.  We estimate that these errors
may also be as large as 30\%.

Figure 4 shows grey-scale images of 5 interlayer vortices in Tl-2201
magnetically imaged using a scanning SQUID microscope\onlinecite{kamsci}.
In these experiments
the experimental signal is equal to the
integrated magnetic flux through the pickup loop.
The vortices appear
elliptical in shape, with the long axis (parallel to the planes) nearly
vertical in these images. The spatial extent of the vortex images
perpendicular to the planes is limited simply by the size of the pickup
loop. The spatial extent parallel to the planes is set primarily by the
interplanar penetration depth.
Figure 5 shows
cross-sections through the experimental
data along the direction parallel to the planes as indicated
by the dashed lines in Figure 4.

To generate a theoretical expression for fitting the experiment
results, we use the full expression (Eq. (\ref{phi})) for the
magnetic fields, taking the a-axis penetration
depth equal to 0.17$\mu$m\onlinecite{zuo} .
An evaluation of Eq.  (\ref{phi}) gives the
$z$-component of the field at a given
height $z_0$ above the sample surface.  This field is then numerically
integrated over the shape of the
pickup loop to obtain the total theoretical flux $\Phi_s(x,y,z)$
through the pickup loop as a function of the
pickup loop position and the c-axis penetration depth $\lambda_c$.
In this case the pickup
loop used was a square 8.2$\mu$m on a side, with a 1$\mu$m linewidth,
and a superconducting shield 5$\mu$m wide which extends to the top
corner of the pickup loop,
as indicated by the inset of
Figure 4. For our modelling we add to the flux through the pickup loop
one third of the flux intercepting an area 5$\mu$m by 5$\mu$m on a side,
starting at the upper corner of the pickup loop,
to account for flux focussing effects from the superconducting shield.
The solid lines in Figure 5 are best fits
of the cross section $\Phi_s(x=0,y,z=z_0)$ to the experimental data,
using the interlayer penetration
$\lambda_c$, the height of the pickup loop $z_0$, and an offset flux,
representing a small uncertainty of the background signal,
as free parameters. The best fit values for each vortex are displayed
in the figure, with uncertainties assigned using a doubling of the
$\chi^2$ value as a criterion.
Reasonable agreement is obtained
between experimental and theoretical cross-sections. The average value
for $\lambda_c$, using a weighting inversely proportional to the square
of the uncertainties, is
$\lambda_c = 18.3\pm3\mu$m.

For comparison, a similar analysis of the same vortices neglecting
the effect of vortex spreading resulted in
a weighted average best-fit value of $\lambda_c = 19\pm2\mu$m. This
comparison is surprising at first, since the
theoretical FWHM at the surface is reduced by 30\% when the vortex
spreading is neglected.  However,
the uncertain value of the height of the pickup loop compensates for
the assumption of negligible vortex
spreading below the surface.
When the spreading below the
surface is neglected, the fitting routine compensates by picking a
higher value of $z_0$, thereby moving
the spreading to the vacuum rather than the superconductor.

In conclusion, we have presented a solution to London's equation
for the case of an interlayer vortex approaching a superconducting
surface normal to the surface and parallel to the planes. This
model is appropriate for experiments which magnetically
image interlayer vortices. Good agreement
with available experiments is obtained with this model,
allowing the quantitative determination of
the interlayer penetration depths from these
measurements.

We would like to thank D.G. Hinks, T.W. Li, and Ming Xu for supplying
the Tl-2201 crystals used for the SQUID images shown in this paper.
We would also like to thank M.B. Ketchen for the design, and M. Bhushan
for the fabrication, of the SQUIDs used here.

\subsection*{Figure Captions}

1. Geometry and axes used in this paper.
A single vortex, centered at $x=0, y=0$, emerges normally to the $ac$
face of the crystal located at $z=0$.

2. Stream-line mapping of $\vec{ {\cal H}}(y,z)$, the magnetic field
integrated over $x$, for a single interlayer vortex shown in Fig. 1. The
spacing of the stream lines is proportional to
$(\partial{\cal H}_z/\partial y)^{-1}$ at $z=-3\lambda_c$.
Fig. 2a is the present model, which includes vortex spreading;
Fig. 2b is for a model which neglects field spreading below the surface.

3. Plot of ${\cal H}_z(y,z)$
for a single interlayer vortex normal to the superconducting surface
$z=0$, as a function of $y/\lambda_c$, for fixed $z$ values,
where $y$ is the distance from the
center of the vortex along a plane direction, and $\lambda_c$ is the
interlayer penetration depth.

4. Grey-scale images of 5 interlayer vortices in the high-T$_c$ cuprate
superconductor Tl-2201. The scaling corresponds to 0.11$\phi_0$(I),
0.16$\phi_0$(II), 0.13$\phi_0$(III), 0.22$\phi_0$(IV), and 0.19$\phi_0$(V)
full-scale variation from black to white in the integrated flux
through the pickup loop.

5. Cross-sections along the plane directions through the images of Figure
4.
Each succesive curve is offset by 0.1 unit for clarity. The dots are the
data; the lines are fits to the present model as described in the text.


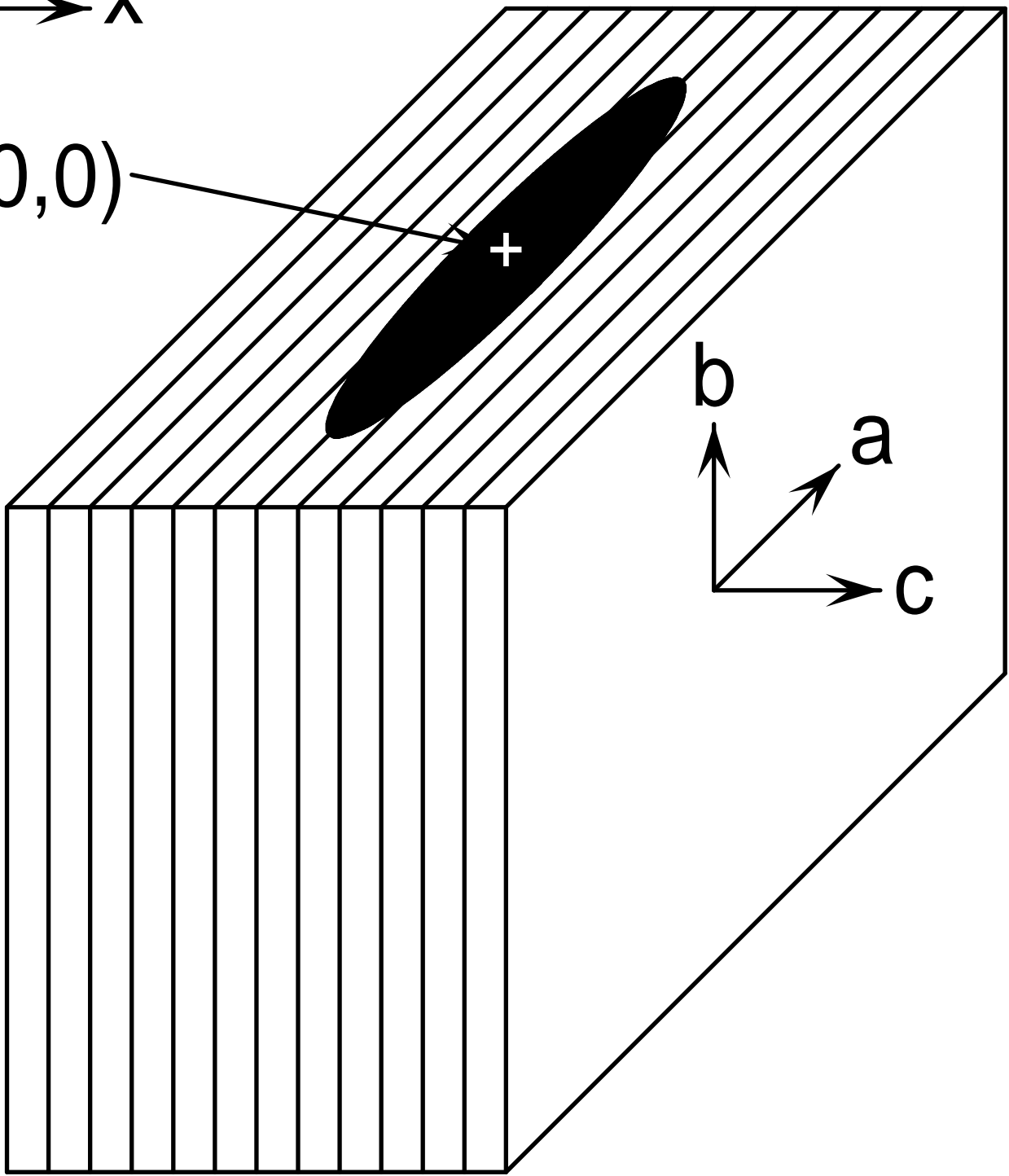
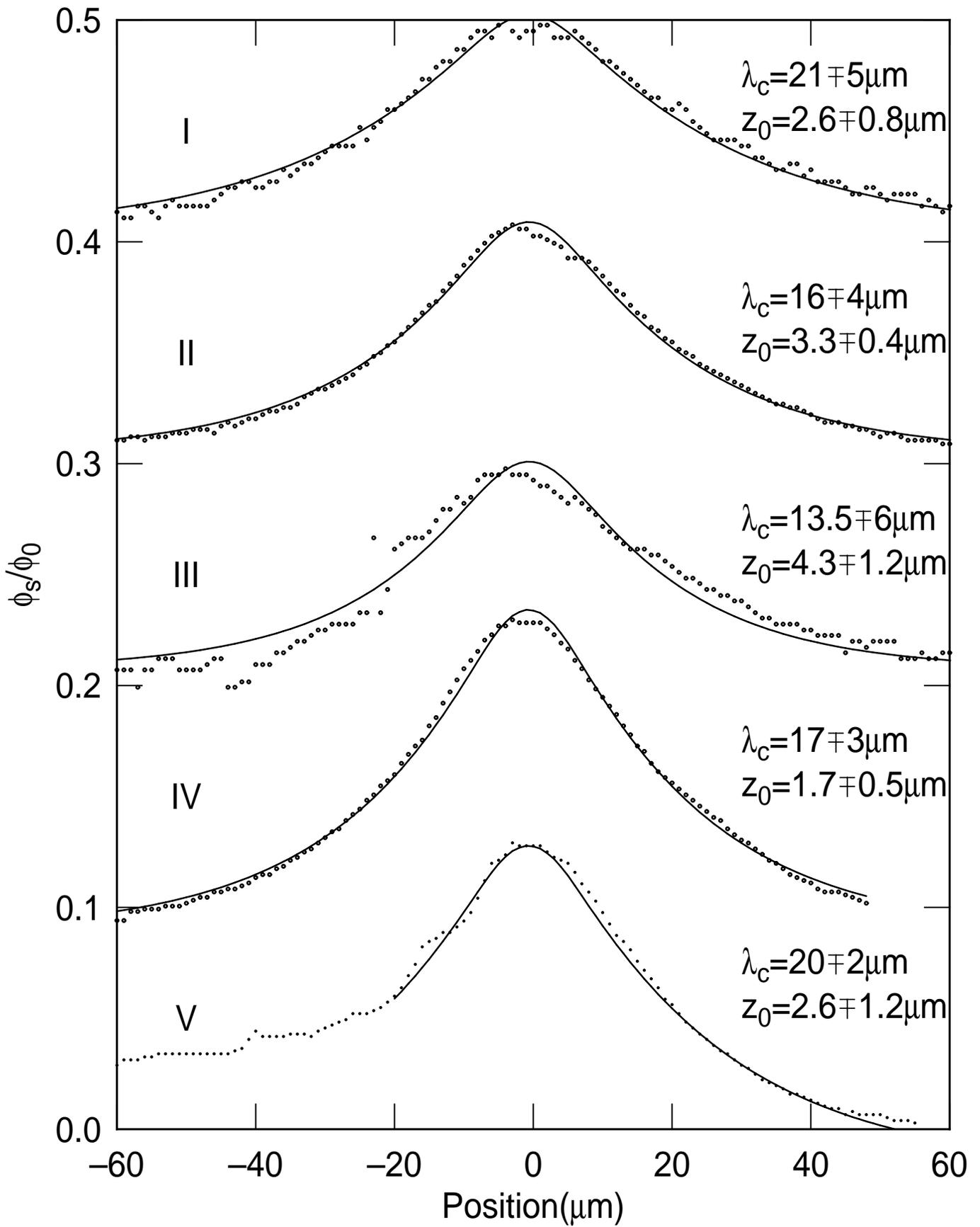
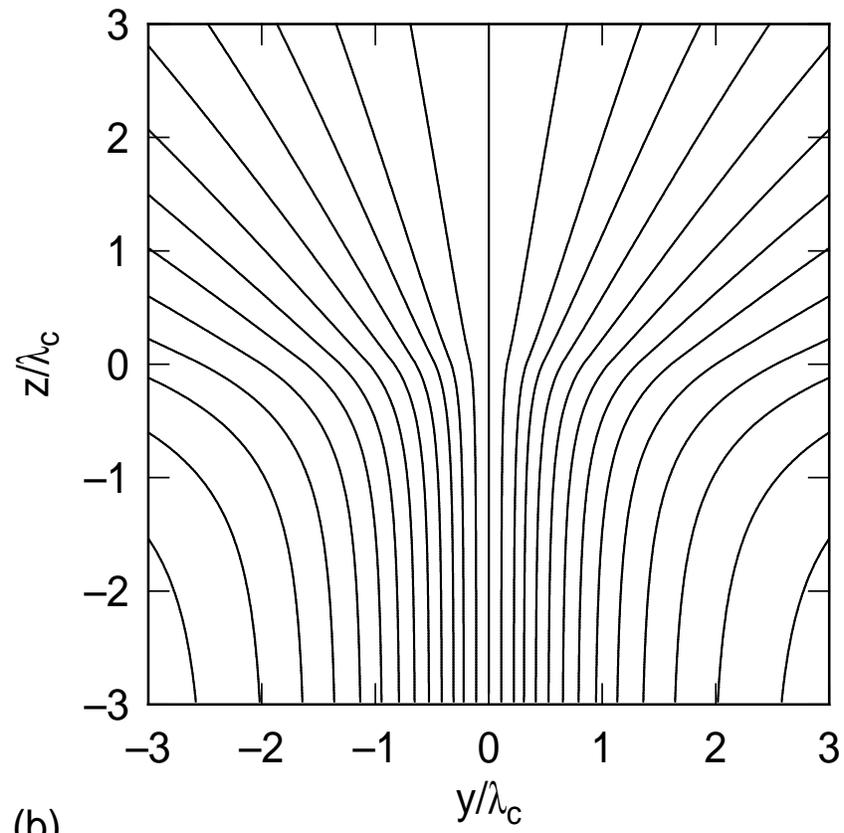
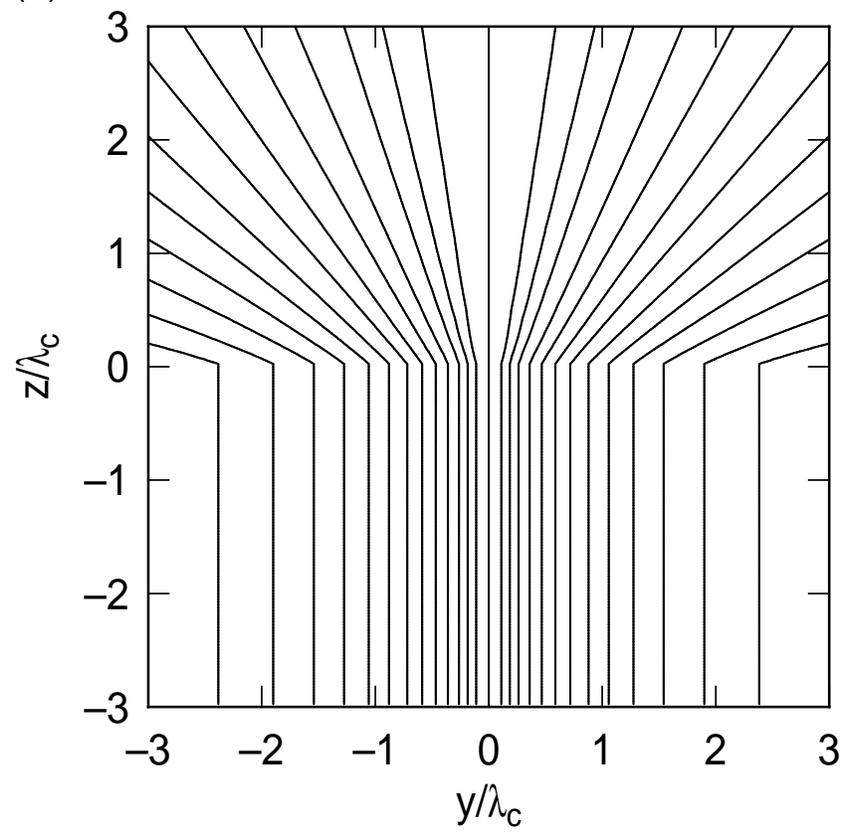
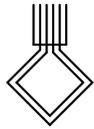
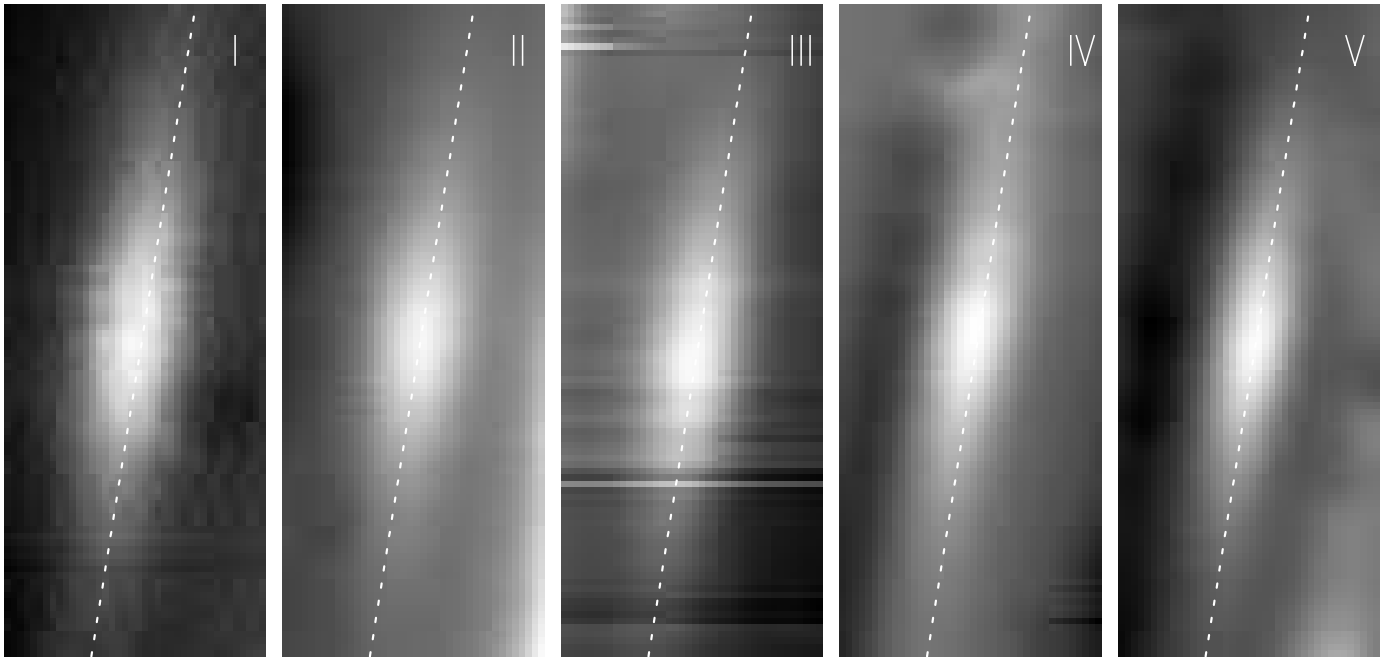
\end{document}